\def\RELEASE{0}  %
\def\ANON{0}     %
\def\SQUEEZE{0}  %

\documentclass[sigconf,10pt]{acmart}
\PassOptionsToPackage{hyphens}{url}\usepackage{hyperref}

\usepackage[noend]{algpseudocode}
\usepackage{algorithmicx}
\usepackage{booktabs}
\usepackage{caption}
\usepackage{comment}
\usepackage[inline]{enumitem}
\usepackage{graphicx}
\usepackage{listings}
\usepackage{multirow}
\usepackage{xcolor}
\usepackage{xspace}
\usepackage{libertine}
\usepackage{microtype}
\usepackage{tikz}
\usepackage{gnuplot-lua-tikz}
\usepackage{tabularx}

\usepackage{subcaption}

\usepackage{relsize} %
\usepackage[most]{tcolorbox}
\tcbuselibrary{xparse,listings}
\NewTCBListing{framedlisting}{O{}}
 {%
  boxrule=0.4pt,
  colback=white,
  colframe=black,
  sharp corners,
  listing only,
  left=0mm, right=0mm, top=0mm, bottom=0mm,
  listing options={#1,frame=none, xleftmargin=0em, framexleftmargin=0em, framextopmargin=0em, belowskip=0em, aboveskip=0em},
 }

\lstdefinestyle{c++}{
    basicstyle=\linespread{0.8}\ttfamily\footnotesize\relsize{-1},
    breakatwhitespace=false,         
    breaklines=true,                 
    captionpos=b,                    
    keepspaces=true,                                     
    numbersep=5pt,                  
    showspaces=false,                
    showstringspaces=false,
    showtabs=false,                  
    tabsize=1,
    xleftmargin=1.5em,
    frame=single,
    framexleftmargin=1.5em,
}

\microtypecontext{spacing=nonfrench}

\hypersetup{
  pdflang={en},
  pdfdisplaydoctitle}

\definecolor[named]{OurPurple}{cmyk}{0.55,1,0,0.15}
\definecolor[named]{OurDarkBlue}{cmyk}{1,0.58,0,0.21}
\hypersetup{
  colorlinks,
  linkcolor=OurPurple,
  citecolor=OurPurple,
  urlcolor=OurDarkBlue,
  filecolor=OurDarkBlue}

\if \SQUEEZE 1
\setlist[itemize]{
  leftmargin=*,
  itemsep=2pt,
  topsep=2pt}

\fi

\def\Snospace~{\S{}}

\newcommand{\fakepara}[1]{\vspace{0mm}\noindent\textbf{#1}\xspace}

\definecolor{mGreen}{rgb}{0,0.6,0}
\definecolor{mGray}{rgb}{0.5,0.5,0.5}
\definecolor{mPurple}{rgb}{0.58,0,0.82}
\definecolor{backgroundColour}{rgb}{0.95,0.95,0.92}

\lstdefinestyle{CStyle}{
    commentstyle=\color{mGreen},
    keywordstyle=\color{magenta},
    numberstyle=\tiny\color{mGray},
    stringstyle=\color{mPurple},
    basicstyle=\footnotesize,
    breakatwhitespace=false,         
    breaklines=true,                 
    captionpos=b,                    
    keepspaces=true,                 
    numbers=left,                    
    numbersep=5pt,                  
    showspaces=false,                
    showstringspaces=false,
    showtabs=false,                  
    tabsize=2,
    language=C
}

\if \RELEASE 1
  \def\NOTES{0}
\else
  \def\NOTES{1}
\fi
\if \NOTES 1
  \newcommand{\XXX}[1]{{\color{red}{XXX {#1}}}}
  \newcommand{\antoine}[1]{{\color{teal}{[\textbf{AK:} {#1}]}}}
  \newcommand{\va}[1]{{\color{violet}{[\textbf{VA:} {#1}]}}}
  \newcommand{\todo}[1]{{\color{blue}{TODO: {#1}}}}
  \newcommand{\fake}[1]{{\color{red}{[\textbf{FAKE:} {#1}]}}}
\else
  \newcommand{\XXX}[1]{}
  \newcommand{\antoine}[1]{}
  \newcommand{\va}[1]{}
  \newcommand{\todo}[1]{}
  \newcommand{\fake}[1]{}
\fi

\newcommand{\eg}{e.g.\xspace}

\if \ANON 1
  \newcommand{\sys}{PrismX\xspace}
\else
  \newcommand{\sys}{Iridescent\xspace}
\fi

\setcopyright{none}
\renewcommand\footnotetextcopyrightpermission[1]{} %

\begin{document}
\date{}
\title{Towards Online Code Specialization of Systems}

\if \ANON 1
  \author{Anonymous Submission \#XXX }
\else
  \author{
    Vaastav Anand,Deepak Garg,Antoine Kaufmann\\
    Max Planck Institute for Software Systems
    }
\fi

\begin{abstract}
  Specializing low-level systems to specifics of the workload
  they serve and platform they are running on often significantly improves
  performance.
  However, specializing systems is difficult because of three
  compounding challenges: i) specialization for optimal performance requires
  in-depth compile-time changes; ii) the right combination of specialization choices
  for optimal performance is hard to predict a priori; and iii) workloads and platform
  details often change online.
  In practice, benefits of specialization are thus not attainable for
  many low-level systems.

  To address this, we advocate for a radically different approach
  for performance-critical low-level systems: designing and
  implementing systems with and for runtime code specialization.
  We leverage just-in-time compilation to change systems code based on
  developer-specified specialization points as the system runs.
  The JIT runtime automatically tries out specialization choices and
  measures their impact on system performance, e.g. request latency or
  throughput, to guide the search.
  With \sys, our early prototype, we demonstrate that online
  specialization
  (i) is feasible even for low-level systems code,  such as network stacks,
  (ii) improves system performance without the need for complex cost
  models,
  (iii) incurs low developer effort, especially compared to manual
  exploration.
  We conclude with future opportunities online system code
  specialization enables.
\end{abstract}

\maketitle

\section{Introduction}

Specializing system implementations to workload
characteristics and hardware can significantly improve
performance and efficiency~\cite{farshin2021packetmill,kalia:erpc,van2017automatic,miano2022domain,sriraman2018mutune,alipourfard2017cherrypick,karthikeyan2023selftune,rzadca2020autopilot,ghasemirahni2024just,wu2024tomur,ghasemirahni2022packet,deng2020redundant,sharif2018trimmer,bhatia2004automatic,molnar2016dataplane,pan2024hoda,sriraman2019softsku}. 
To achieve these benefits for particular hardware and workload combination,
system developers manually modify the system code and recompile
to benefit from the compile-time optimizations enabled by
specialization~\cite{wintermeyer2020p2go,panchenko2019bolt}.

System specialization comes at the cost of generality.%
A system heavily specialized to workload and hardware either performs
poorly outside of this regime, or completely fails.
As a result, developers today must carefully navigate this
specialization-generalization tradeoff and optimize for the most
common hardware and workload setting.

However, even selecting this design point a priori is difficult for multiple reasons.
First, the workload and platform are often unknown at development time.
Systems are often deployed to different hardware than they were
developed on.
Second, workload and platform often change at runtime.
A system might migrate to newer deployments over time.
Application workloads rarely remain fixed throughout the
application lifetime.
Third, even for a known workload and platform, performance for the
modified system is hard to predict.
Modern system performance is dictated by emergent behaviors
arising from the complex intra-system interactions and
interactions between the system and the workload.
To make matters worse, the exact same system and workload are likely
to exhibit different performance on different hardware.

As a result, making optimal system code specialization choices
is fundamentally an iterative process: run the system, profile, adjust
the code, and repeat.
Unfortunately this is also extremely laborious for developers, and
thus rarely worth the cost and complexity in practice.

We propose a fundamentally different approach to system specialization
for performance and efficiency: \emph{automated online
specialization}.
We aim to specialize systems at runtime to the exact workload 
and hardware conditions where the system is deployed, \emph{without
the human developer in the loop}.
To demonstrate the idea's feasibility, we present a design
proposal for a specialization toolbox that can iteratively specialize
low-level systems online with runtime performance feedback based on
developer-annotated possible \emph{specialization points}.
We have developed an initial prototype \sys, and provide early
evidence for feasibility, performance improvements, and lowered
developer effort.
Further, we argue that this mechanism of code changes at runtime with
full system performance could fundamentally change how we build
systems and how we approach compiler optimization.
\section{Specialization is Effective but Complex}%
\label{sec:bg}

We illustrate the potential and pitfalls specializations for
optimizing low-level system code with a well understood algorithm:
blocked/tiled matrix multiplication to optimize cache locality~\cite{carr1989blocking,wolfe1987iteration,wolfe1989more}.
By dividing up input and output matrices into blocks and operating on
these instead of full columns or rows, cache locality improves
substantially.
We measure different combinations of (square) input matrix sizes and
block sizes 5 different processor core architectures
(see rows and columns in \autoref{tab:mmul_hw}).
Blocking works as expected. 
We find that blocking reduces required processor cycles by $1.15\times$
to $28.5\times$ with the ideal block size.

\begin{table}[t]
\centering
\begin{tabular}{lccc}
\toprule
  \textbf{Machine / Workload} & \textbf{N=1024} & \textbf{N=25}6
  &\textbf{N=64} \\ 
\midrule
IceLake    & 32     & 32    & 32   \\
IvyBridge         & 16     & 16    & 4    \\
CoffeeLake       & 32     & 4     & 4    \\
AlderLake-p & 32     & 4     & 2    \\
AlderLake-e & 64     & 4     & 4    \\
\bottomrule
\end{tabular}
\caption{Optimal configurations for our block matrix
  multiply, across 5 hardware platforms and 3 workloads.}
\label{tab:mmul_hw}
\vspace{-8mm}
\end{table}

\fakepara{Specialization improves performance.}
When trying to implement this in a system we need to choose a block
size to use.
Unfortunately we find that the optimal block size depends on the
hardware the system is running on, different cache sizes or latencies,
and the workload,
such as different matrix sizes.
We illustrate this in \autoref{tab:mmul_hw} where we show the best
performing block size for each combination, and find 5 different
optimal block sizes depending on the datapoint.
Thus choosing a specific block size is specializing the implementation
to the machine and workload, reduces performance in other cases.

\fakepara{Code changes are necessary.}
This is a challenge since we find that the block size really has to be
fixed at compile time.
Because of the computationally dense nature of the algorithm, good
peformance really requires heavy compiler optimizations, such as
loop unrolling and vectorization.
And since blocking dictates the trip count of the two innermost loops,
these optimizations are not feasible if the block size is a variable.
Here we find that using a variable instead of a fixed block size can
reduce performance by up to $6.5\times$.

\fakepara{Optimal choices are hard to predict.}
Our results in \autoref{tab:mmul_hw} show that the optimal
parameters are difficult to predict a priori, even for a fixed
workload.
Depending on the machine three different block sizes may be ideal.
With dynamic workloads that change over time, or external factors such
as noisy neighbor cores filling up caches, this quickly becomes
completely unpredictable. This combination leaves the system developer with no good options:
The best performance requires compile-time specialization, but the
best choices are dynamic and unpredictable.
The closest solution for simple cases is to implement
multiple different versions of the code and switch between them at
runtime.
However, this is laborious and does not scale beyond a single
parameter.
\section{Specializing without Human in-the Loop}
\label{sec:approach}

We enable system developers to easily
build systems that perform well across a broad set of workloads and
platforms with a fundamentally different approach.
Instead of exclusively compiling ahead of time, we argue that this
requires \emph{online specialization} through code
changes as the system runs.
This enables one version of the system to automatically adapt to the
workload and hardware conditions.
Specializing online makes it possible to directly use system
performance as feedback for choosing optimal configurations.

We propose to use just in time (JIT) compilation to be able to change
the code as it runs.
We combine this with developer provided annotation for potentially
relevant specialization points in the code.
Finally we leverage overall system performance measurements, such as
request throughput, to guide an automated, empirical search for the
best configuration.

\fakepara{Lightweight developer instrumentation is sufficient.}
Finding useful specialization opportunities in low-level systems code
is extremely challenging for compilers.
However, developers have excellent intuition about \emph{possible
specialization points}.
We thus require developers to provide lightweight annotations about
potential values in the code that could be specialized to compile time
constants.
The key insight here, is that commonly simply turning a variable in
the code into a constant, then enables a host of cascading compiler
optimizations through constant propagation, dead code, elimination,
unrolling, etc.
Thus we can leverage minimal annotation to automatically make
substantial changes to the code.

\fakepara{Fallbacks for correctness are necessary.}
When automatically specializing and since real systems are complex,
incorrect specializations are inevitable.
Crashes or incorrect behavior for these cases are not a viable
outcome.
To avoid this, we include guards in the specialized code to make sure
the conditions hold.
If a guard triggers, we fall back to the original code version for
this execution, incurring a performance penalty but no other problems.

\fakepara{Online measurements can replace cost models.}
The combination of JIT and specialization points provides the
necessary mechanisms to specialize code at runtime, but leaves the
runtime with a large number of potential knobs to turn.
Ahead of time compilers making similar decisions automatically
typically rely on built-in cost models.
Unfortunately, cost models for system performance are inevitably
highly inaccurate and would result in suboptimal results.
However, since we are making these decisions online as the system
runs, we can simply do what human developers do as well: try out
different options and measure their impact on full system performance.
Online specialization thus enables fully accurate decisions for
specialization choices based on measurements rather than models.

\section{An Online Specialization Runtime}
\label{sec:design}

\begin{figure}[t]%
\centering%
\includegraphics[width=\linewidth]{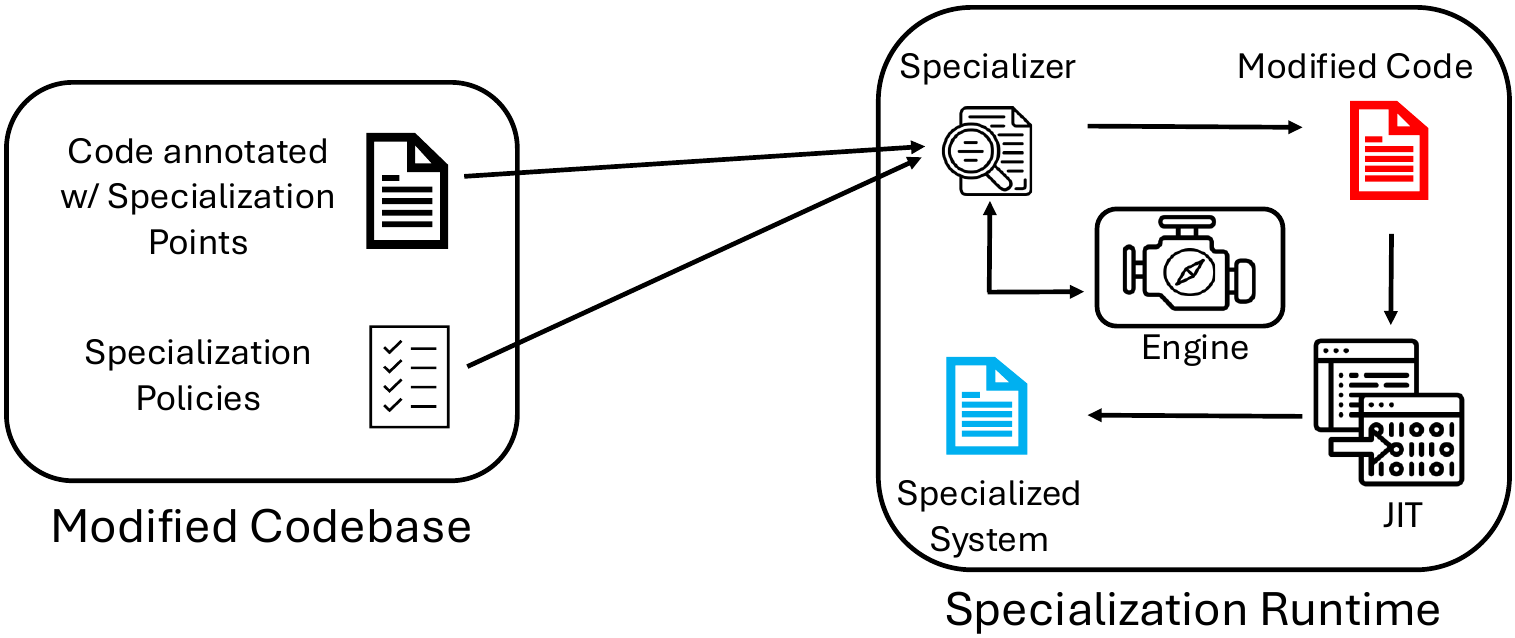}%
\caption{\sys design}%
\label{fig:sys_design}%
\vspace{-3mm}%
\end{figure}

We present a design proposal for enabling online specialization.
\autoref{fig:sys_design} shows the design of \sys, a runtime specialization toolbox that allows
developers to automatically specialize any given software based
on metrics calculated at runtime. \sys supports exploration of the
set of all possible specializations, applying a given particular specialization,
and switching between different specializations at runtime to handle
changing workloads.

\subsection{System Modifications}

\fakepara{Source-Code Division.} Developers divide the codebase into two parts: (i) the performance-critical \emph{handler code}; (ii) the \emph{fixed code}.
The handler code is the set of functions and their dependencies in the codebase that the developer wishes
to specialize. In contrast, the fixed code entails the general-purpose part of the codebase that makes function calls
into the handler code to execute incoming requests.
For example, in the MMulBlockBench microbenchmark, \texttt{\footnotesize{matrix\_multiply}} would comprise the performance-critical handler code.
The callers of this function would comprise the fixed code.

\fakepara{Specialization points.} The developer instruments the handler code in the specialization runtime to create
\emph{specialization points}. A specialization point is a variable or a parameter in
the handler code that can potentially be converted into a constant during execution.
Broadly, there are two types of specialization points: (i) workload knobs; and (ii) configuration knobs.
Workload knobs are variables that change with potentially every request in
a given workload. For example, in the MMulBlockBench microbenchmark, the block size $s$ is the workload knob.
In contrast, configuration knobs are variables that do not rely on workload data
and are usually provided as input to the program as part of a command line flag or a configuration file.
For example, this would be the batch size of a packet processing framework.

\fakepara{Specialization Policies.} Through the specialization policies, developers control the exploration and selection of specializations.
The developers initialize the specialization runtime with the target end-to-end metric and a way to measure the metric
and the specific exploration policy to use to explore the specialization space of the handler code.

\subsection{Specialization Runtime}

To specialize the handler code, the developer first places
the handler code in the \emph{specialization runtime}. The specialization runtime is a wrapper
around a JIT that manages the lifetime of the handler code inside the JIT.  

\fakepara{Specializer.} To specialize a point with a given value, the specializer specializes
the handler code by replacing the variable in the code that is mapped to the specialization point and fixes
its value to the specific value. The handler code is then recompiled in the JIT compiler by the specialized runtime.
The recompilation process allows for various compiler optimization passes to take effect that now view
the previously runtime specialization point as a fixed compile time constant. This allows for the optimization
passes to be more aggressive with optimizations such as constant propagation, loop unrolling, instruction
vectorization, among others to produce a specialized version of the handler code.

Even though the specialization points are applied at a variable level, the code specialization
takes place at the granularity of a function. That is, once a specialization point is specialized, 
a new version of the function containing that specialization point will be generated
with the variable's value now fixed to the specialized value.

\begin{table}[t]
    \begin{tabularx}{\linewidth}{|l|X|}
    \hline
    Hook                       & Description                              \\ \hline \hline
    \small{runtime\_init}              & Initialize runtime                       \\ 
    \small{update\_runtime}            & Recompile handler code               \\ 
    \small{get\_specialized\_function} & Get address of specialized function      \\ 
    \small{point\_specialize}          & Specialize point to a value            \\ 
    \small{point\_disable\_spec}              & Disable specialization       \\ 
    \small{point\_disable\_spec\_check}       & Disable specialization check \\ 
    \small{point\_enable\_collection}         & Enable data collection       \\ 
    \small{point\_disable\_collection}        & Disable data collection      \\ 
    \small{start\_exploration}         & Start exploration              \\ 
    \small{end\_exploration}           & End exploration                \\ \hline
    \end{tabularx}
    \caption{Runtime hooks provided by \sys}
    \label{tab:hooks}
    \vspace{-3mm}
\end{table}

\fakepara{Handling errors and side-effects.} To ensure correctness,
for each specialized function, the specialization runtime inserts a runtime check called a specialization check that checks
if an input value matches the current specialized value of the runtime. This check allows for the runtime
to detect if a new incoming request in the workload meets the criteria to use the specialized version of the code.
If this check fails, then the specialization runtime cleans up any side-effects and transfers control back
to the generic version of the function. To ensure correct clean up of side effects, the developers define and
register a dedicated cleanup function in the handler code for each function that could be specialized by the runtime.
It is critical to note here that not all side-effects are reversible (\eg sending a packet to a neighbor), so \sys
performs a best-effort clean-up.

\fakepara{Using the specialized code}. To use the specialized version of the handler, the fixed code must obtain the addresses of all possible
specialized functions from the specialization runtime. The fixed code must use these function addresses to make function calls
to the specialized version of the handler code. If no specialization has been enabled for any given specialization
point, then the function address defaults to the non-customized generic version of the function. 

\fakepara{Specialization Hooks.} The specialization runtime provides hooks in the fixed code that the developers can use to control the behavior
of the specialization runtime.
\autoref{tab:hooks} provides a list of basic operations that are supported by the runtime.
The operations include which
specialization point(s) to specialize, when to specialize them, when to change the specialization, and updating the objective function to use for selecting the optimal configuration.
Developers can use the hooks in the fixed code to change the behavior of the runtime.

\fakepara{Exploration Engine.} As the space of possible specializations and their combinations is large,
it is almost impossible for developers to know which possible specializations must be enabled in conjunction
for best performance. Thus, the fixed code needs a way to
try out different specializations and then settle on a set of specializations.

To achieve this, \sys provides an explorer engine that can be configured 
to explore different specializations for the various specialization points. 
The runtime instruments the handler code to start collecting three pieces
of information for the specialization point - (i) the set of values for that point, (ii) the frequency with which each value
is seen in the workload, and (iii) the number of specialization check failures.
The developer can override or turn-off the data collection process by providing a fixed set of values
that should be used during the exploration.

\subsection{Prototype Details}

We have implemented our current prototype in $3K$ lines of C++ code. Our current prototype uses the JIT compiler provided by LLVM~\cite{lattner:lllvm}
to generate the specialized code at runtime. The specializer is implemented as a set of LLVM transformation passes that operate on LLVM IR of the handler code.
We are also in the process of implementing a prototype of online specialization in Go and integrating it into the microservice generation framework, Blueprint~\cite{anand2023blueprint}.

\section{Use Cases}
\label{sec:case}

We showcase how we can use online specialization with \sys for three use-cases: (i) enabling compile-time optimizations at runtime;
(ii) enabling incremental specializations at runtime; (iii) design exploration at runtime.

\subsection{Compiler Optimizations at Runtime}

\begin{table}[t]
\centering
\begin{tabular}{lccc}
\toprule
\textbf{Machine}          & \textbf{Constant} (c) & \textbf{Variable} (v) & \textbf{Benefit} \\ 
(Processor)      & (cycles/op)  & (cycles/op) & (v/c)\\
\midrule
IceLake    & 175297    & 284944             & 61\% \\ 
IvyBridge         & 250434    &   661295           & 246\% \\ 
CoffeeLake       & 168817    &  581130             & 348\% \\ 
AlderLake-p & 173350    & 583724             & 336\% \\
AlderLake-e & 206924    & 557572           & 269\% \\
\bottomrule
\end{tabular}
\caption{Impact of turning $s$ as compile-time constant at runtime for N=64 for different hardwares (\autoref{tab:mmul_hw})}
\label{tab:mmul_compilation}
\vspace{-3mm}
\end{table}

Compile-time optimizations often produce more performant code.
These compile time optimizations include
constant propagation, loop unrolling, dead-code elimination, and use of vectorization instructions.

With \sys, we can enable compile time optimizations at runtime. 
To show the potential benefits of enabling compile-time optimizations at runtime, we use the MMulBlockBench
shown in \autoref{tab:mmul_hw}. Specifically, for the workload $N=64$, we compare the performance of keeping the optimal block size,$s$,
as a runtime parameter to that of converting it into a constant at runtime using \sys. For both configurations,
we execute the function for a fixed number of times and measure the amount of cycles spent for executing 1 execution of the function.
\autoref{tab:mmul_compilation} shows the benefit for converting $s$ into a compile-time constant for different hardwares.
For each hardware, we get at least 50\% reduction in consumed cycles, and greater than 240\% reduction in consumed cycles for four of the five operating conditions.

\subsection{Incremental Specialization}

Incremental computing and has been a mature idea~\cite{ramalingam1993categorized} for decades.
In recent years, incremental specialization has been used for packet processing frameworks~\cite{ruffy2024incremental,miano2022domain,farshin2021packetmill}.
To show that \sys can easily enable incremental specialization, we extend the Network Functions from Vigor~\cite{zaostrovnykh2019verifying} and Pix~\cite{iyer2022performance}. Specifically, we target the lpm (longest prefix match) network function, which matches the incoming packet's destination
address to one of its rules in its prefix table and then routes the packet based on the selected rule.
We extend this function with \sys to enable a specialization similar to the hot map specialization of Morpheus~\cite{miano2022domain}.
In this specialization, \sys divides the execution into a monitoring phase and a specialization phase. 
In the monitoring phase, \sys initially instruments the lpm code to find the most common input addresses and store the calculated routes for these addresses.
After a user-specified amount of time, \sys converts the collected information into hard-coded values at the entry point of the lpm function
to return the hard-coded route if the input address is one of the hardcoded addresses.
With \sys, we see a 9\% increase in the throughput.

\subsection{Runtime Design Exploration}

Automatically searching over a design space at runtime has been successfully applied for
optimizing platform configuration settings~\cite{sriraman2019softsku,somashekar2024oppertune},
in our use-case we target design space exploration of the deployed system itself.

\begin{figure}[t]%
\centering%
\includegraphics[width=\linewidth]{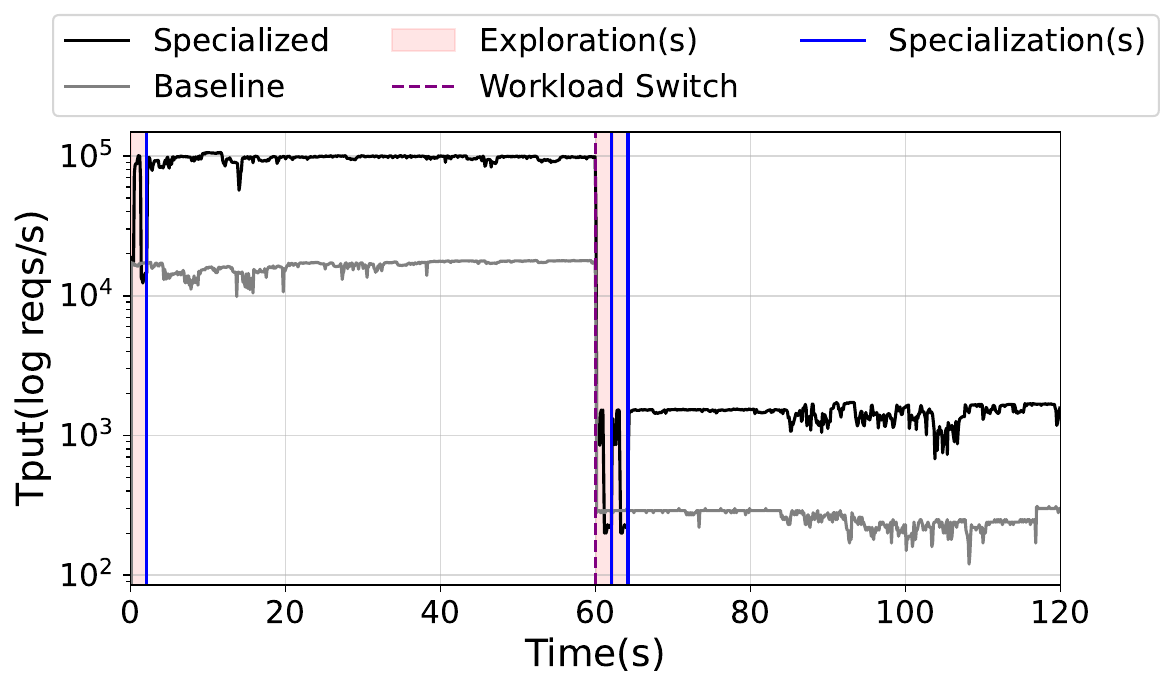}%
\caption{Automatic Exploration and Specialization of MMulBlockBench for two different workloads}%
\label{fig:mmul_adaptation}%
\end{figure}

To show that \sys can enable automatic runtime design exploration, we first use the MMulBlockBench microbenchmark.
We execute a \sys-enabled version of the function and a non-\sys enabled baseline version.
In the non-baseline version, the block size $s$ is a runtime parameter. In the \sys-enabled version, \sys explores different
values of $s$ by specializing the value in the code and converting $s$ into a compile-time constant.
We execute each version with two different workloads in succession, each lasting one minute.
We measure the overall throughput (executions/s) for both configurations.
\autoref{fig:mmul_adaptation} shows how
\sys can easily explore the different block sizes and find
a performant configuration as compared to the baseline.
Moreover, \sys can automatically detect the workload change based on the
drop in throughput and restart the A/B testing process.

\begin{figure}[t]%
\centering%
\includegraphics[width=\linewidth]{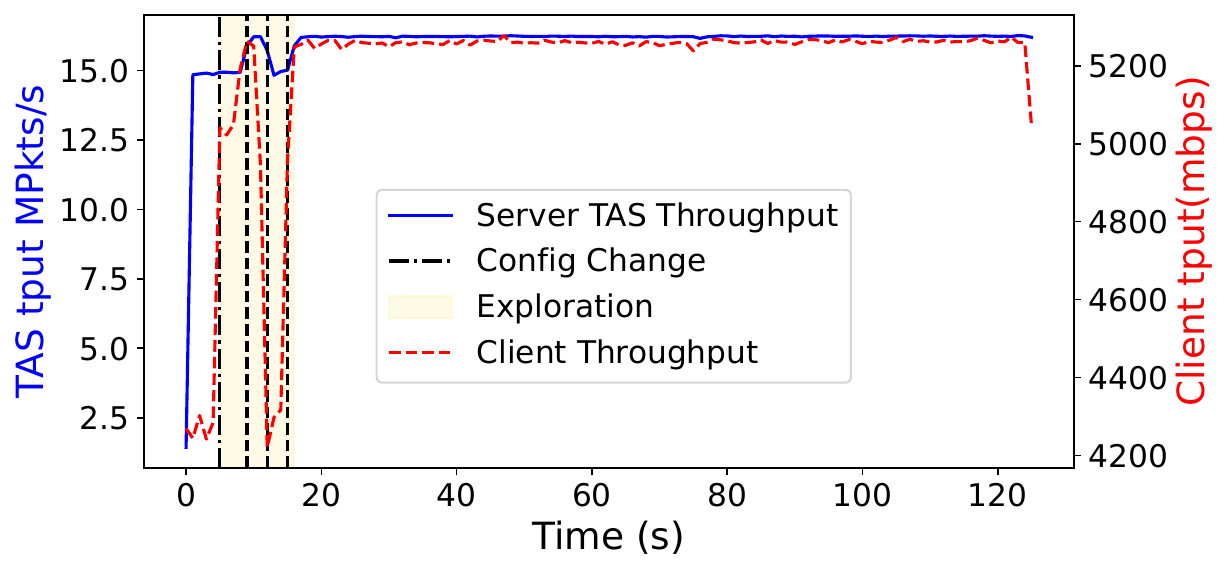}%
\caption{Automatic Exploration and Specialization of different configurations for TAS}%
\label{fig:tas_adaptation}%
\end{figure}

Next, we show runtime design exploration for TAS~\cite{kaufmann:tas}, a TCP acceleration stack.
We setup our experiment with a typical server-client setting, with the server
providing an echo service. The client is multi-threaded and executes an open-loop
workload of 64 byte packets. Both the server and client are enabled with TAS.
In our experiment, we modify the server-side TAS and measure the throughput of the
server side TAS as millions of packets processed per second. 
With \sys we explore different values of the batch size in the TAS source code on the server side.
Normally, the batch size is a fixed constant throughout the TAS source code. 
Instead of treating it as a homogeneous constant, we convert batch size into three different variables
for three different locations in the source code and explore different combinations of the values of the batch sizes.
\autoref{fig:tas_adaptation} shows how \sys can automatically select the best-configuration on the server side
by automatically exploring the different values of the batch size for different locations in the source code.
\section{Outlook}
\label{sec:future}

We believe that online optimization of systems with metric-guided exploration can open multiple research avenues.
We highlight some promising future directions below.

\fakepara{Empirically enabled Compiler Optimizations.} In our demonstration, we have only focused on one aspect of specialization - converting runtime parameters
into compile time constants. Developers can develop custom compile-time
optimization techniques that are designed specifically for leveraging the structure of their system and combining it with empirical information
to optimize the system.
For example, offline profile-guided prefetching instruction injection LLVM pass of APT-GET~\cite{jamilan2022apt} 
can seamlessly be integrated with \sys to better utilize the prefetching abilities of the processor at runtime.
Moreover, incorporating domain-specific techniques like offline compiler optimizations of PacketMill~\cite{farshin2021packetmill}
into \sys represent interesting future specialization opportunities.

\fakepara{Leveraging scale \& ML for exploration.}
Several ML-based techniques exist for tuning configurations~\cite{van2017automatic,akgun2023improving,cereda2021cgptuner,li2018metis,lin2022adaptive,mahgoub2020optimuscloud,somashekar2022reducing,somashekar2024oppertune,duan2009tuning,jamilan2022apt,swaminathan2017off,ruffy2024incremental}. 
OPPerTune~\cite{somashekar2024oppertune} provides access to a plethora
of ML techniques toolbox for automatically searching through large configuration spaces. We believe \sys can be integrated with OPPerTune to utilize state-of-the-art exploration techniques. We believe developers can leverage the large-scale nature of modern cloud systems and explore the large specialization spaces
by measuring the performance of different specializations in parallel.

\fakepara{Automatic Online System Repair.} \sys can be extended with techniques to automatically detect performance problems and
then automatically explore the design space of the system to fix the issue.
For example, \sys can be used to detect and fix emergent misbehaviors~\cite{mogul2006emergent} caused by complex interactions
between different components of the system.
\section{Conclusion}

We introduced online specialization to enable automatic online adaptation to workload and hardware specifics. 
With our proposed design of \sys, we enable online optimization of performance-critical systems by dynamically exploring and applying the most effective specialization.
\sys enables developers to potentially achieve significant performance improvements without the complexity of manual specialization. 
We believe, \sys can be used as a building block for enabling custom specializations at runtime to extract maximum performance
from systems. \if \ANON 0
\fi

\bibliographystyle{plain}
\bibliography{paper,bibdb/papers,bibdb/strings,bibdb/defs}

\label{page:last}
\end{document}